\documentclass[]{article}

\usepackage{graphicx}
\usepackage{float}
\usepackage{url}
\usepackage{authblk}
\title{\textbf{pySiDR: Python Event Reconstruction for SiD} \\ \vspace{0.25in} \large \textit{Talk presented at the International Workshop on Future Linear Colliders (LCWS2019), Sendai, Japan, 28 October-1 November, 2019. C19-10-28.}}
\author[]{C.T. Potter}
\affil[]{Center for High Energy Physics, University of Oregon }

\date{\today}

\begin{document}

\twocolumn

\maketitle

\begin{abstract}
Event reconstruction in the ILC community has typically relied on algorithms implemented in C++, a fast compiled language. However, the Python package pyLCIO provides a full interface to tracker and calorimeter hits stored in LCIO files, opening up the possibility to implement reconstruction algorithms in a language uniquely well suited to working with large lists of hits built with list comprehensions. Python, an interpreted language which can perform complex tasks with minimal code, also allows seamless integration with powerful machine learning tools developed recently. We discuss pySiDR, a Python package for SiD event reconstruction. 
\end{abstract}

\section{Introduction}

The Silicon Detector (SiD) \cite{Aihara:2009ad,Behnke:2013lya} is one of two technically mature detectors designed for the International Linear Collider (ILC) \cite{Behnke:2013xla,Baer:2013cma,Phinney:2007gp}, a proposed next generation $e^+ e^-$ collider. Event reconstruction of simulated data in the ILC detector community relies on algorithms implemented in C++ and integrated with the executable \texttt{Marlin} in the ILCsoft framework \cite{ilcsoft}. The compiled code is high level C++ which can be inaccessible to a nonexpert and challenging to develop on CVMFS \cite{cvmfs}, sprawling over thousands of lines in numerous distinct files. 

\begin{figure}[]
\includegraphics[width=0.48\textwidth]{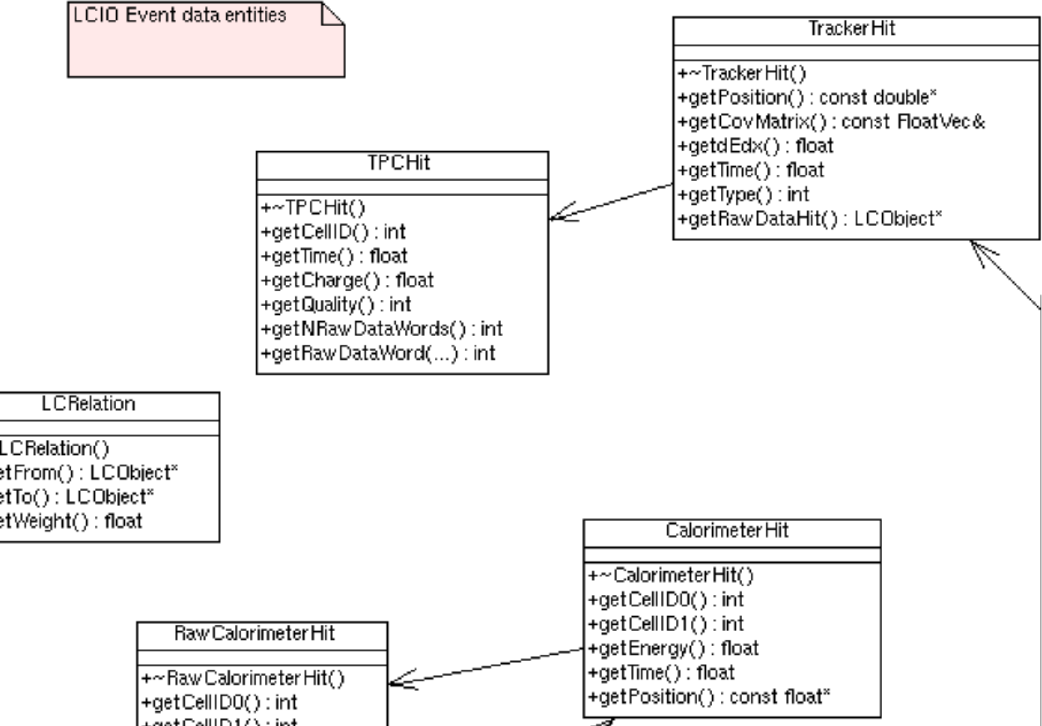}
\caption{Methods available in LCIO for obtaining tracking and calorimeter hit information. Note in particular the \texttt{getPosition()}, \texttt{getEnergy()} and \texttt{getTime()} methods.}
\label{fig:lcio}
\end{figure}

The package pyLCIO \cite{pylcio} provides complete access to tracker and calorimeter hits in LCIO \cite{gaede2003lcio,gaede_2017} files and therefore allows the possibility to implement event reconstruction algorithms in Python. Python code is interpreted, giving immediate feedback to developers and users. It is terse, centralized in a few files, self documenting and easily understood by a wide range of users. When compiled to bytecode the timing is competitive. Finally, the interface to TensorFlow \cite{tensorflow2015-whitepaper}, Scikit-Learn \cite{scikit-learn} and other machine learning tools is smooth. See Figure \ref{fig:lcio} for the LCIO methods for accessing detector hits.

\begin{figure*}[th!]
\hspace{0.4in}\includegraphics[angle=90,width=0.4\textwidth]{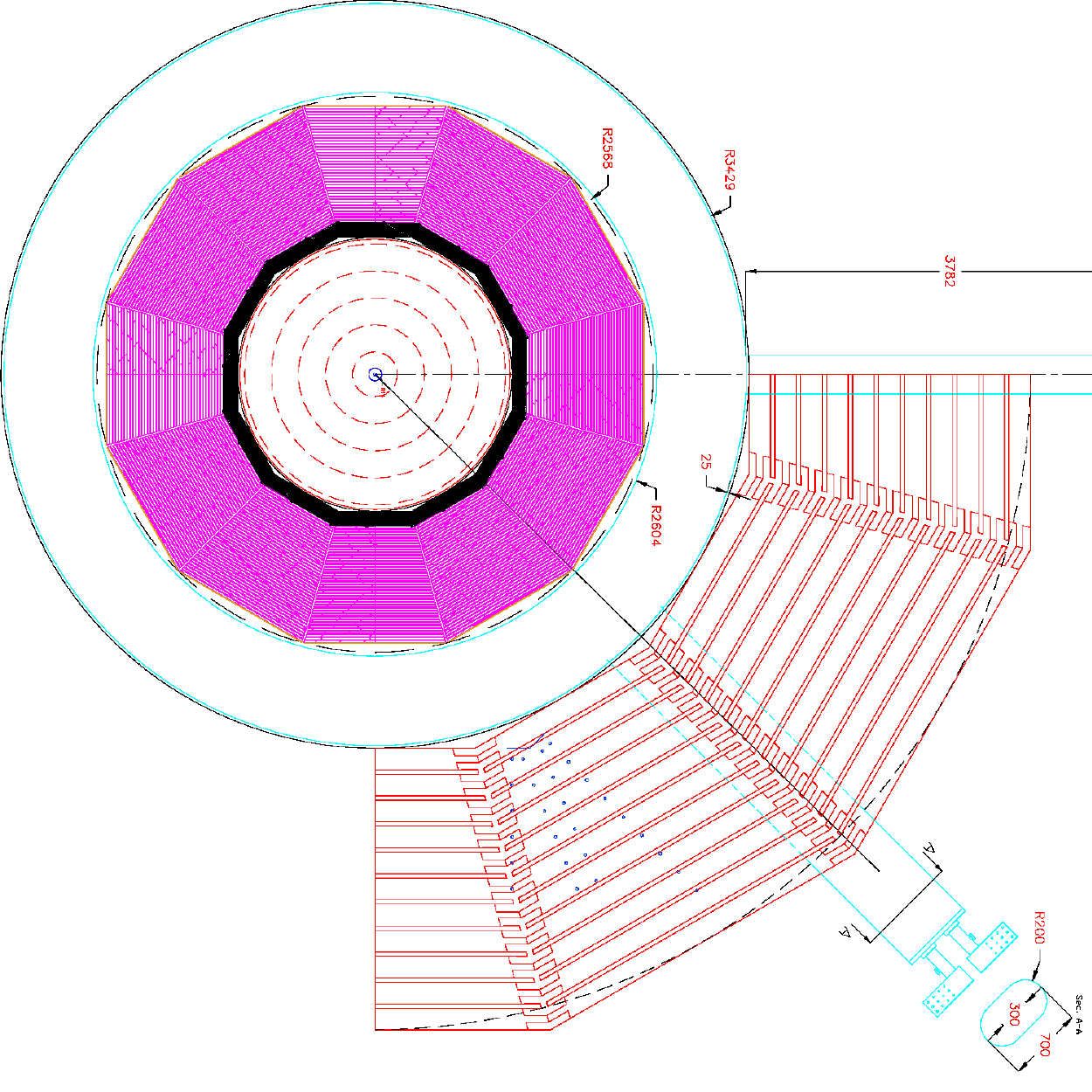}
\hspace{0.4in}\includegraphics[width=0.4\textwidth]{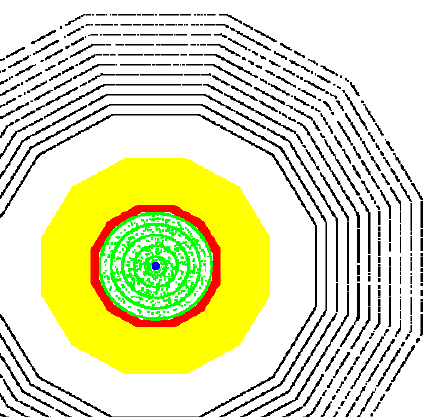}
\caption{At left, technical drawing of SiD barrel view. At right, $(x,y)$ hit maps from $10^3$ Higgs boson events $e^+ e^- \rightarrow ZH$ with $Z\rightarrow \mu^+ \mu^-$. Modeled using the SiD compact detector description in DD4hep with ILCsoft v02-00-02.}
\label{fig:sid}
\end{figure*}

In this work we describe the pySiDR \cite{sidperformance} package, a package for ILC event reconstruction with LCIO files. While pySiDR was written for SiD, it is almost totally detector \emph{geometry agnostic} and works with other detector designs with minimal modification. 

The SiD detector comprises a precise Silicon Vertex Detector and Tracker, together with an electromagnetic calorimeter made of alternating layers of Tungsten absorber and Silicon sensitive elements, and a hadronic calorimeter (HCal) with Resistive Plate Chamber sensitive elements. A 5T solenoid provides the necessary magnetic field to enable precision particle flow throughout the calorimetry. A muon detector is instrumented in the steel flux return.  See Figure \ref{fig:sid} for a technical drawing of the SiD detector and hit maps generated from Higgs boson events in ILCsoft v02-00-02 using the compact SiD detector description.

\section{Technicalities}

For single particle events, we use the LCIO particle gun in lcgeo/example from GitHub to generate a flat distribution of single electrons, photons, charged pions, neutrons and muons with energy $1 < E < 125$~GeV. For Higgstrahlung events, we assume $\sqrt{s}=250$~GeV. We use Whizard 2.6.3 with polarized beams and initial state radiation. Beamstrahlung is included using Guineapig 1.4.4 and the staged ILC250 beam parameters. Full simulation of these events is performed with DD4hep and the compact SiD\_o2\_v03 detector description in ILCsoft v02-00-02.

We use Python 2.7.10, the default for ILCsoft v02-00-02, and the Python package pySiDR, compiled into bytecode, from SiDPerformance on GitHub. The pySiDR package contains three files:

\begin{itemize}
\item pySiDR.py: track fitting , track and cluster finding, particle flow
\item utilities.py: often used mathematical functions, linear and circular regression 
\item sid\_o2\_v3.py: SiD detector geometry file
\end{itemize}

\noindent The geometry file included is detector specific for SiD, but any other detector geometry file is easily adapted.

In the following section we focus on the track and cluster finding as well as the particle flow algorithm.

\section{Algorithms}

\subsection{Track Finding}

\begin{figure*}[t]
\includegraphics[width=\textwidth]{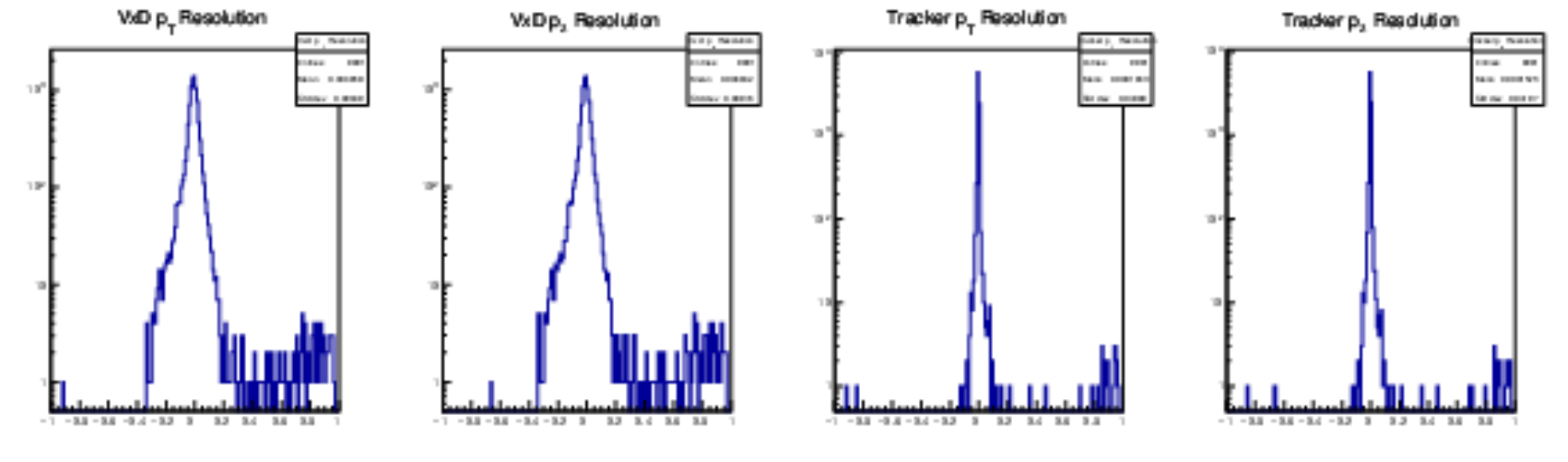}
\includegraphics[width=\textwidth]{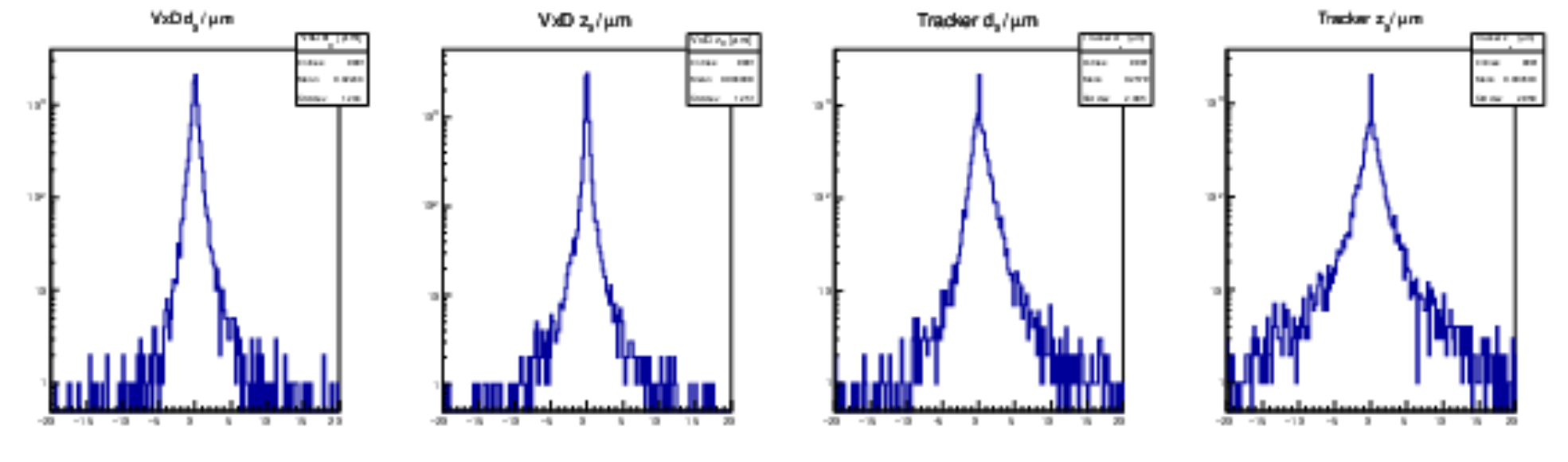}
\caption{Track $p_T$ and $p_z$ resolution (top) and impact parameter $d_0$ and $z_0$ (bottom) for the Vertex Detector (left) and Tracker (right) obtained with single muon events with $1 < E < 125$~GeV. We obtain $\Delta p/p \approx 0.09$ (0.04) and $d_0,z_0 \approx 1\mu$m (3$\mu$m) for the Vertex Detector (Tracker).}
\label{fig:tracks}
\end{figure*}

Track finding begins with seeds formed from Vertex Detector hits. All possible doublets of hits are combined with an ersatz hit $(0,0,0)$ at the interaction point to form hit triplets. The two real hits are taken from Vertex Detector layers specified by the user. The seeds are then fit with a circle in $(x,y)$ and a line in $(s,z)$, where $s$ is arclength, and the track is extrapolated throughout the Vertex Detector and Tracker. Any barrel (endcap) hit in any layer satisfying a maximum distance criterion after track extraplolation to the transverse (azimuthal) distance of the hit is associated to the track.

Thus charged particles which fail to leave a hit in one or more tracking layers can be recovered by requesting seeds from all possible combinations of two layers in the Vertex Detector. Seeds may also be taken from the Tracker and extrapolated inward to the Vertex Detector. Duplicate tracks which share a maximum number of hits can be eliminated with a dedicated clone removal function. 

Track parameters are determined after accumulation of hits, with the ersatz hit $(0,0,0)$ removed, from the Vertex Detector and Tracker. A minimum number of hits and a maximum $\chi^2$ are imposed to remove fake tracks. With single muon events, we obtain $\Delta p/p\approx 0.09$ (0.04) and $d_0,z_0 \approx 1\mu$m (3$\mu$m) for the Vertex Detector (Tracker). See Figure \ref{fig:tracks} for the $p_T$ and $p_z$ resolution and transverse $d_0$ and azimuthal $z_0$ impact parameters. 

\subsection{Cluster Finding}

\begin{figure*}[t]
\includegraphics[width=\textwidth]{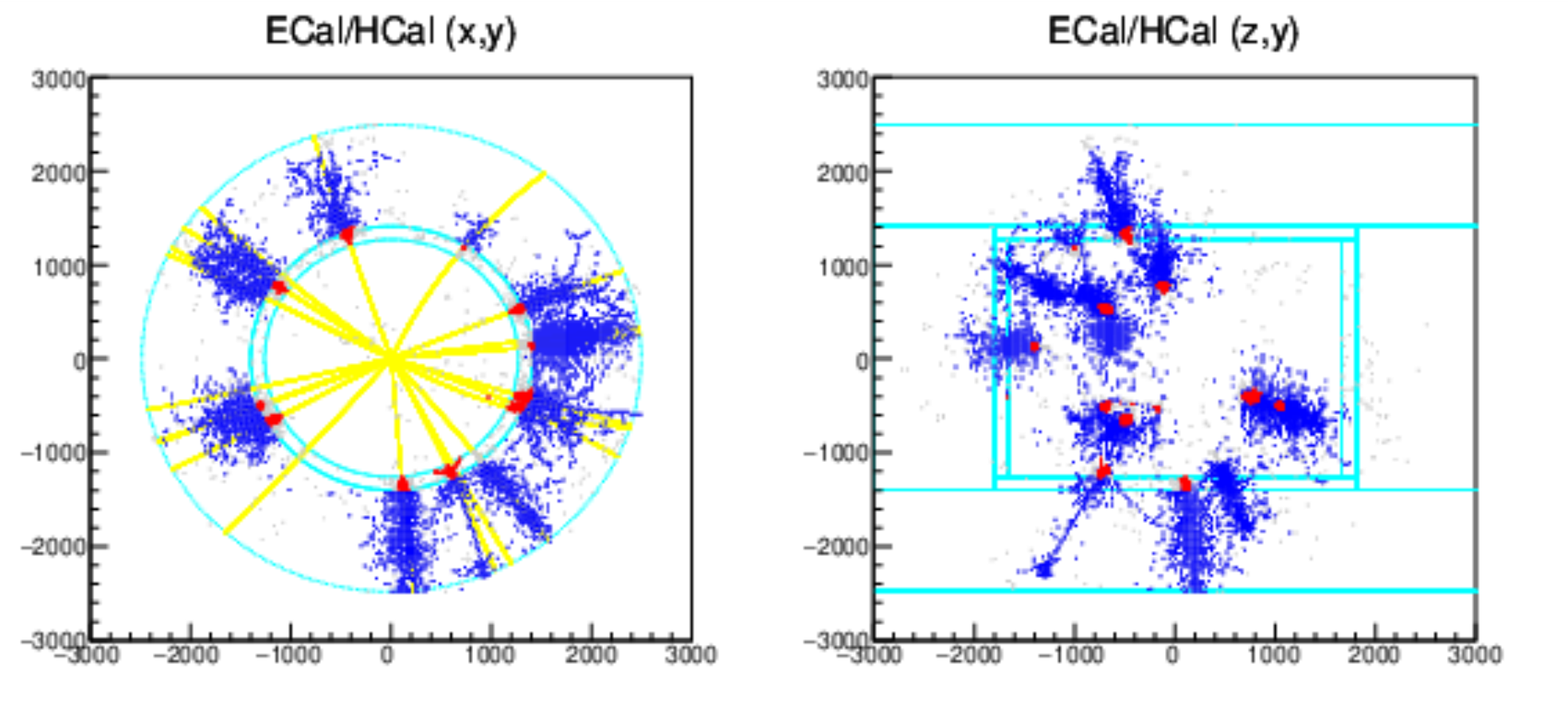}
\caption{Superimposed single charged pion events with unclustered calorimeter hits (grey), clustered ECal hits (red), clustered HCal hits (blue) and the Monte Carlo truth trajectories (yellow). The $(x,y)$ view is on the left, while the $(z,y)$ view is on the right.}
\label{fig:clusters}
\end{figure*}

The default cluster finder in pySiDR is a topological cluster finder. Three hit energy thresholds are defined by the user: seed threshold, add cell threshold and neighbor threshold. First all calorimeter hits with energy larger than the seed threshold form seeds for hit accumulation. Then any adjacent hits satisfying a maximum distance criterion with energy larger than the add hit threshold are accumulated to the hit, and any adjacent hit with energy larger than the neighbor threshold forms a seed for recursive hit accumulation. Any clusters which share hits are merged. 

ECal and HCal cluster finding are performed with the same algorithm but different maximum distance criteria, each based on calorimeter cell and layer geometry.  See Figure \ref{fig:clusters} for hits and calorimeter clusters created by single charged pion events.

One application of machine learning techniques with Python to SiD calorimetry, predicting calorimeter leakage energy, is described in Section \ref{sec:nn}.

\subsection{Particle Flow}

Particle flow, a concept central to the design of SiD, is implemented in with a straightforward extrapolation of tracks to calorimetry. Each track identified by the track finding is extrapolated to the transerse (azimuthal) distance of the barrel (endcap) cluster centroid and the track is associated to the cluster which minimizes the azimuthal (transverse) distance. After all tracks have been associated to clusters, any remaining clusters are considered to be made by electrically neutral particles. 

If a track associates to both an ECal and an HCal cluster, the track is considered an electron if the associated ECal cluster energy is larger than the associated HCal cluster energy, and a charged hadron otherwise. If an ECal cluster unassociated to a track is adjacent to an HCal cluster, the neutral is considered a photon if the ECal cluster energy is larger than the adjacent HCal cluster energy, and a neutral hadron otherwise. Thus lists of electrons, photons, charged hadrons and neutral hadrons are formed.

In the case of charged particles, the particle flow technique allows the association of calorimeter clusters to tracks, thus replacing and inherently imprecise cluster energy measurement with a precise track momentum measurement. The precision of the cluster energy measurement is particularly poor when the shower starts late in the ECal or HCal and some energy leaks beyond the calorimeter containment. In contrast to charged particles, for clusters created by neutral particles there is no leakage remediation from particle flow. In the following section we discuss a machine learning technique for recovering this calorimeter energy leakage.

\subsection{Energy Leakage \label{sec:nn}}

\begin{figure*}[t]
\includegraphics[width=0.5\textwidth]{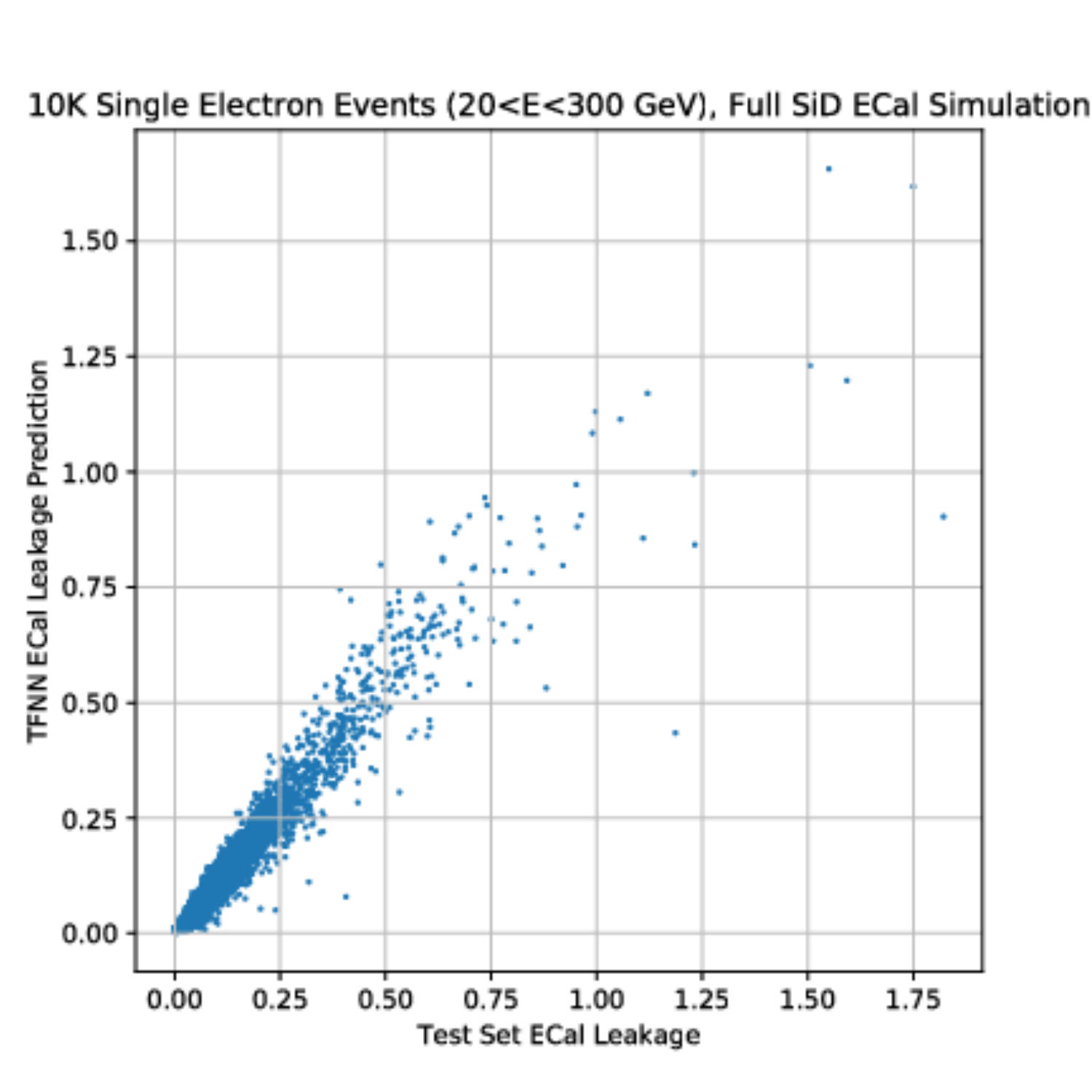}
\includegraphics[width=0.5\textwidth]{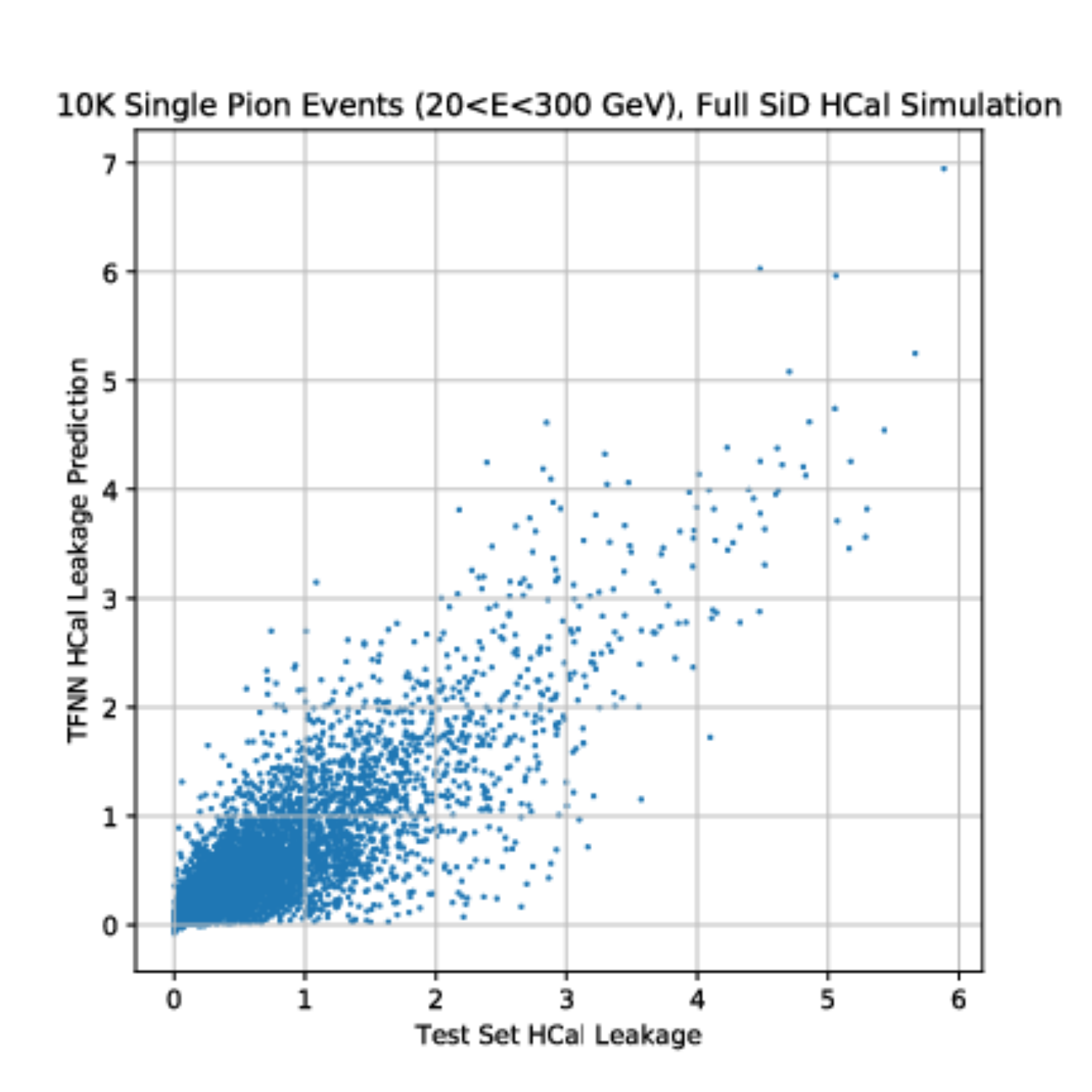}
\caption{Predicted leakage energy vs actual leakage energy in the SiD ECal (left) for $10^4$ single electron events and HCal (right) for $10^4$ single charged pion events. For the HCal neural network, the precision is expected to improve when energy deposits from the ECal are included.}
\label{fig:leakage}
\end{figure*}

One example of the application of powerful machine learning tools to event reconstruction is illustrated by a TensorFlow neural network designed to predict the leakage energy in the SiD calorimetry based on the shower energy profile. While the SiD ECal (HCal) is 26$X_0$ ($4.5\lambda$) deep, showering is a statistical process, so some showers develop late and leak energy from the back of the calorimetry. This energy leakage limits the calorimeter energy resolution.

However, this leakage energy can be recovered with a neural network which predicts the leakage energy based on the input energy deposits in each calorimeter layer. See Figure \ref{fig:leakage}. It should be noted that, while the precision of the HCal energy prediction is expected to be less precise than the ECal precision due to the physical nature of the showering processes, the ECal energy deposition has not yet been included in the HCal neural network. We expect that when the ECal information is included in the HCal neural network, so that the full shower profile is sampled, the prediction precision will improve.

For conference proceedings related to this study of machine learning in calorimeter leakage recovery and resulting performance improvement, see \cite{me} and \cite{masako}.

\section{Timing}

Python is an interpreted language, which accounts for its flexibility and power, but when uncompiled it slows execution time to an unacceptable level. Fortunately, when imported Python compiles to bytecode, an intermediate language which executes far faster than uncompiled Python. While bytecode will not always run faster than other compiled languages, the inherent advantages of Python make it competitive.

First we consider the timing performance of the executable \texttt{Marlin}, the reconstruction program in ILCsoft. With Marlin, the reconstruction sequence is defined in an XML file with \emph{processors}. The Conformal Tracking and PandoraPFA processors are invoked in the execute tag of the reconstruction steering file input to \texttt{Marlin}. 
Next we consider the timing performance of pySiDR compiled to bytecode. Here the clustering is performed as a distinct stage in reconstruction, in contrast to the case with \texttt{Marlin}, where the clustering is performed by PandoraPFA. See Table \ref{tab:timing} for the timing results with \texttt{Marlin} and pySiDR run on the same $10^4$ single charged pion events on the same computer using one 3.4GHz core.

\begin{table}[t]
\begin{center}
\begin{tabular}{|l|c|c|} \hline
 & \texttt{Marlin} & pySiDR \\ \hline 
Tracks &  70ms (64\%) &   8.9ms(29\%) \\ \hline
+Clusters & - &  30ms(68\%) \\ \hline
+PFlow & 110ms (36\%) & 31ms(3\%) \\ \hline
\end{tabular}
\caption{Cumulative time per event for \texttt{Marlin} and pySiDR to run over single charged pion events obtained with the Linux \texttt{time} command. In \texttt{Marlin} cluster finding is performed in the PandoraPFA particle flow  processor.}
\label{tab:timing}
\end{center}
\end{table}

\section{Conclusion}

We have described pySiDR, a Python package for reconstructing objects in simulated SiD events. Traditionally detector digitization and event reconstruction for ILC detectors has been implemented in compiled C++, which runs on events stored in LCIO files after detector simulation. An alternative model combines detector simulation and digitization in the same step, with event reconstruction in Python. 

The package can be found in SiDPerformance in ILCsoft on GitHub. pySiDR is a work on progress, not a polished final product, and can easily be improved. Expected future development includes leakage recovery, bremstrahlung recovery, jetfinding and vertexing. Despite its name, pySiDR is largely geometry agnostic and can be easily adapted to other detector designs.

\bibliography{paper}

\end{document}